\documentstyle[preprint,eqsecnum,aps,tighten]{revtex}

\begin{document}

\title{Comprehensive analysis of conditionally exactly solvable models}

\author{Rajkumar Roychoudhury and Pinaki Roy}

\address{Physics and Applied Mathematics Unit, Indian Statistical
Institute, \\
Calcutta 700035, India }

\author{Miloslav Znojil}

\address{\'{U}stav jadern\'e fyziky AV \v{C}R, 250 68 \v{R}e\v{z},
Czech Republic,\\ electronic address: znojil@ujf.cas.cz}

\author{G\'eza L\'evai}

\address{Institute of Nuclear Research of the Hungarian Academy of
Sciences, \\
Pf. 51, Debrecen, 4001 Hungary}

\maketitle

\begin{abstract}
We study a quantum mechanical potential introduced previously as a
conditionally
exactly solvable (CES) model. Besides an analysis following its original
introduction in terms of the point canonical transformation, we also
present an alternative supersymmetric construction of it. We demonstrate
that from the three roots of the implicit cubic equation defining the
bound-state energy eigenvalues, there is always only one that leads to
a meaningful physical state. Finally we demonstrate that the present
CES interaction is, in fact, an exactly solvable Natanzon-class
potential.
\end{abstract}

\vspace{5mm}


\newpage

\section{Introduction}

Exactly solvable models have attracted much attention since the
early years of quantum mechanics. Some solvable potentials have become
standard examples of text books, but a lot more have been discovered
by various approaches. Systematic work has been done to generate and
classify these potentials using the factorization method \cite{fact},
algebraic methods \cite{alg} and more recently, in terms of
supersymmetric quantum mechanics (SUSYQM) \cite{Witten}. These approaches
were found to be interrelated with each other \cite{ms89,lg89,lg94}.

The most general family of solvable potentials is the six-parameter
Natanzon class \cite{nat}, which contains potentials with solutions
expressible in terms of a single (confluent) hypergeometric function.
A rather important subclass of this is that of the shape-invariant
potentials \cite{si}, to which the most well-known potentials (such as
the harmonic oscillator, Coulomb, P\"oschl--Teller, etc.) potentials
belong. Altogether 12 such potentials have been identified
\cite{lg89,Khare}, but some of these, actually represent different
forms of the same potentials, and their separate discussion is justified
only for historical reasons. An important recent development was the
introduction of supersymmetric quantum mechanics (SUSYQM), which
can be considered a re-interpretation of the factorization method
\cite{fact}, and which links basically isospectral potentials in a
pairwise manner. Shape-invariant potentials are defined
in terms of SUSYQM: the functional form of the SUSYQM partner
potentials has to be the same, and only the parameters appearing in
them can be different.

SUSYQM has been found rather useful in generating new solvable
potentials as SUSYQM partners from known solvable ones. A rather
wide potential class is obtained as the SUSYQM partner of Natanzon
potentials, but these are not Natanzon potentials themselves (except in
the case of shape-invariance), since their solution is written as the
linear combination of several (confluent) hypergeometric functions
\cite{cks87}. There are also further solvable potentials which are
solved by functions other than the (confluent) hypergeometric type.
Examples for this are the square well \cite{Flugge} and the exponential
potential, which are solved by Bessel functions.

A different concept of solvability characterizes quasi-exactly solvable
(QES) potentials \cite{Ushveridze}. In this case only part of the
eigenstates
can be obtained, by requiring termination of a recursion relation
defining the eigenfunctions in a polynomial form.

The most recent concept of solvability is related to conditionally
exactly solvable (CES) potentials. The first models coined CES
potentials \cite{Dutra,Dutt} were characterized by the fact that
the coupling constant of some potential term had to be fixed to a
numerical constant value in order to obtain their solutions.
These potentials were introduced by the point canonical transformation
method \cite{pct}. Here we present the analysis of one of these CES
potentials \cite{Dutt}. Our motivation is to clarify some
inconsistencies in their treatment, and to determine their place
in other classification schemes of solvable models.
(We note that another class of CES potentials was also
introduced using the techniques of SUSYQM \cite{susyces,pla},
but we do not extend our analysis on this class.)

In Sec. \ref{ni} we give a re-interpretation of the potential of Ref.
\cite{Dutt} in a supersymmetric context, and derive the bound-state
energies determined implicitly by a cubic equation. In Sec. \ref{san}
the procedure is placed in a more general context of methods based on
variable transformations, and the potential is identified as an exactly
solvable member of the Natanzon potential class.

\section{The model of Dutt, Khare and Varshni }
\label{ni}

We start with presenting the potentials introduced by Dutt et al.
\cite{Dutt} as CES models. The two potentials defined on the
full axis $x \in (-\infty,\infty)$ can be written in a common form as
\begin{equation}
 V^{(g_0,g_1,g_2,g_3)}(x)=
 {g_0
 \over
 e^x\,
 z(x)
 }+
 {g_1
 \over
 z^{}(x)
 }+ {g_2
 \over
 z^{2}(x) }+ {g_3
 \over
 z^{4}(x)
 }\ ,
\label{vgs}
\end{equation}
with $ z(x) = (1+e^{-2x})^{1/2} \in (1,\infty)$.
The explicit form of these potentials \cite{Dutt} is
 \begin{equation}
 V_1^{(DKV)}(x) = V^{(0,-B,A,-3/4)}(x),
 \ \ \ \ \ \ \
  V_2^{(DKV)}(x) = V^{(-B,0,A,-3/4)}(x)\ .
  \label{trip}
 \end{equation}
These potentials depend on two parameters ($A$ and $B$) which define
the potential shape. The coupling constant of the third potential term
has to be fixed to a constant value ($-3/4$) in order to obtain exact
solution of these models. This is why the authors of Ref. \cite{Dutt}
identified these potentials as CES ones.

One can easily demonstrate that the two potentials, in fact, are
equivalent in the sense that
 \begin{equation}
 V^{(0,-B,A,-3/4)}(x) =
 V^{(-D,0,C,-3/4)}(-x)  + \varepsilon\ ,
   \label{tripolis}
  \end{equation}
where
\begin{equation}
 \varepsilon = -A + 3/4,
 \ \ \ \ \ \
 C = -A+3/2, \ \ \ \ \ \
 D = B\ .
\label{vdkvrel}
\end{equation}
Thus, in what follows it is sufficient to deal with only one of the
potentials, so we pick $V_1^{(DKV)}(x)$ for our analysis.

\subsection{Conventional approach via the point canonical transformation}
\label{ni-ichi}

In Ref. \cite{Dutt} potentials (\ref{trip}) were introduced using the
point canonical transformation method \cite{pct}, by which a
Schr\"odinger-type differential equation can be transformed into
another equation of this type, applying an invertible parametrization
$r=r(x)$. With this change of variables, dating back to Liouville
\cite{Liouville} a given asymptotically free equation
 \begin{equation}
\left[-\,\frac{d^2}{dr^2} + U(r)\right]\, \chi(r) = -\kappa^2
\,\chi(r)
 \label{SEor}
  \end{equation}
can be transformed
into an apparently different bound state problem
 \begin{equation}
\left[-\,\frac{d^2}{dx^2} + V(x)\right]\, \psi(x) = -k^2\,\psi(x).
 \label{SEder}
  \label{SE}
  \end{equation}

After we denote the derivative by a prime ($x'(r)$ etc.), an
extremely elementary correspondence between the potentials and/or
energies is obtained,
 \begin{equation}
 U(r)+\kappa^2=\left [
 x'(r)
 \right ]^2
 \left \{
 V[x(r)]+k^2
 \right \}
+
 \left (
 \frac{3}{4}
 {x''(r) \over x'(r)}
 \right )^2
 -
 \frac{1}{2}
 {x'''(r) \over x'(r)}\ .
 \label{newpot}
  \end{equation}
Obviously, the ``old'' energy eigenvalues are related to the
parameters of the ``new'' potential, and vice versa.
The formal definition of the new wave functions is also virtually
trivial,
  \begin{equation}
 \psi(x)
 = \left(x'[r(x)]\right)^{1/2} \chi[r(x)]\ .
  \label{trenky}
  \end{equation}
In any situation of practical interest one may just pick up a
suitable exactly solvable (ES) problem (\ref{SEor}) and derive quickly
its partner (\ref{SEder}). Setting out from two
shape-invariant \cite{si} ES potentials defined on the positive half axis,
Dutt et al. \cite{Dutt} used the variable transformation $x=\ln(\sinh r)$
to obtain potentials (\ref{trip}). The particular initial potentials
and their energies were
 \begin{equation}
 U_{1}(r) = -2b\frac{\cosh\,r}{\sinh\,r}
 +a(a-1)\,\frac{1}{\sinh^2r},
 \ \ \ \
 \kappa^2=\kappa^2_m=(a+n)^2+b^2/(a+n)^2
 \label{u1pot}
 \end{equation}
[with $b > (a+n_{\rm max})^2$] and
 \begin{equation}
 U_{2}(r) = -(2a+1)b\ \frac{\cosh\,r}{\sinh^2r}
 +\left [a(a+1)+b^2 \right ]\,\frac{1}{\sinh^2r},
 \ \ \ \
 \kappa^2=\kappa^2_n=(a-n)^2
 \label{u2pot}
 \end{equation}
(with $b > a > n_{\rm max}$).

Recalling the bound-state wave functions of potentials $U_j(r)$,
the solutions to potentials $V_j^{(DKV)}(x)$ in (\ref{trip}) readily
follow from Eq. (\ref{trenky}). Without the loss of generality
we can consider the $j=1$ case and recall the solutions of $U_1(r)$
(see e.g. \cite{Khare,lg89}) in terms of Jacobi polynomials,
 \begin{equation}
 \chi(z) =
 (z-1)^{-\frac{1}{2}(a+n-s)}
 (z+1)^{-\frac{1}{2}(a+n+s)}
 P_n^{(-a-n+s,-a-n-s )}
 (z), \ \ \ \ s = b/(a+n)
 \label{u1wf}
 \end{equation}
with $z=z(r)=\coth r$.
Using this function in (\ref{trenky}), substituting it into the
Schr\"odinger equation and matching parameters $a$ and $b$ with $A$
and $B$ appearing in $V_1^{(DKV)}(x)$ in (\ref{trip}), we find
$B=2b$ and
\begin{equation}
 A=n^2+1/2 +(2n+1)\,a +b^2/(a+n)^2 .
 \label{eu}
\end{equation}
This equation will ultimately determine the energy eigenvalues of
quantum number $n$, through a cubic equation as described also in
Ref. \cite{Dutt}. We postpone the analysis of this formula to Sec.
\ref{ni-san}, where our new results concerning the energy spectrum
of the $V_j^{(DKV)}(x)$ potentials are presented. Before that, we
present an alternative interpretation of the same problem in terms
of a supersymmetric framework.

\subsection{Supersymmetric construction}
\label{ni-ni}

An interesting SUSY re-interpretation of the solvability of
Schr\"{o}dinger equations has been described by Nag et al.
\cite{Nag}. They have employed the two Dutra's models \cite{Dutra}
in order to
illustrate their main idea. Unfortunately, the spectrum of states
in the latter potentials can only be determined purely numerically
\cite{Stillinger}. Strictly speaking, the potentials do not belong
to the CES class \cite{comment,reply}. At best, only their
incomplete ( = quasi-exact) non-numerical solution can be obtained
at certain {\em exceptional} energies and couplings
\cite{comment,Classif}. Within the SUSY methodical framework, they
seem less suitable for illustrative purposes.

We shall now obtain the spectrum of the potential $V_1^{(DKV)}(x)$ in
(\ref{trip}) in a manifestly supersymmetric fashion. Before doing
this we recall that in supersymmetric quantum mechanics \cite{Khare}
a pair of Hamiltonians $H_{\pm}$ defined by
\begin{equation}
H_{\pm} = -\frac{{\rm d}^2}{{\rm d}x^2} + V_{\pm}(x)
        = -\frac{{\rm d}^2}{{\rm d}x^2}+ W^2(x)\pm W'(x)
\end{equation}
are isospectral except for the zero energy ground state, which, for
unbroken supersymmetry exists only for one of the partner potentials,
$V_-(x)$. The ground-state solution of $H_-$ is related to the $W(x)$
superpotential through
\begin{equation}
W(x)=-\frac{\rm d}{{\rm d}x}\ln \psi_0^{(-)}(x)\ .
\label{sspot}
\end{equation}
One can also extend the concept of superpotential to the excited
states of $V_-(x)$, simply using $\psi_n^{(-)}(x)$ in (\ref{sspot}).
In this case $W(x)$ has singularities at the nodes of $\psi_n^{(-)}(x)$,
and one can talk about singular superpotentials \cite{Khare}.
(Note that such singularities cannot occur using the nodeless
ground-state wave function $\psi_0^{(-)}(x)$.) Despite these singularities
of $W(x)$, it can be shown \cite{Khare} that $V_-(x)$ will be
singularity-free in this case too, and these will appear only for
the partner potential $V_+(x)$. Our purpose is, however, to discuss
only $V_-(x)$, which we identify with $V_1^{(DKV)}(x)$ in Eq.
(\ref{trip}), in a supersymmetric form, therefore we shall
avoid the problems arising due to the singularities of $W(x)$.

For this purpose, let us consider the superpotential
\begin{equation}
W(z) = \frac{B_1}{z} - \frac{1}{2z^2} - C_0 + \sum_{i=1}^n
\frac{g_i^{'}(z)}{g_i(z)} \ ,
\label{wy}
\end{equation}
where $z=(1+{\rm e}^{-2x})^{1/2}$ as in (\ref{trip}) and (\ref{vgs}),
and $g_i(z)$ is given by
\begin{equation}
g_i(z) = \frac{1}{1+g_i z}~~~,~~~C_0 = \epsilon_0^{1/2} \ ,
\label{c0}
\end{equation}
where $\epsilon_n$ is related to the (negative) bound-state energies of
potential $V_1^{(DKV)}(x)$ via $\epsilon_n=-E_n$.
Note that the zero-energy wave function
$\psi_0^{(-)}(x) = N_0 \exp[-\int W(x) dx]$ is always normalizable
for our choice of $W(x)$, irrespective of the values of $g_i$.
It may be noted that if we had omitted the last term in (\ref{wy}) i.e.,
\begin{equation}
W_0(z) = \frac{B_1}{z} - \frac{1}{2z^2} - C_0 \label{w0}
\end{equation}
we
would have obtained only the ground state. Insertion of the last term
containing the sum ensures that we would get the excited states also.

It is straightforward to show that $W(z)$ can be written in the form
\begin{equation}
W(z) = \frac{B_1-\sum_{i=1}^{n} g_i}{z} - \frac{1}{2z^2} + C_0^{'} +
\frac{\sum_{i=1}^{n}(g_i^2-1)}{(1+g_iz)} \ ,
\label{Wy1}
\end{equation}
where we have defined $C_0^{'} = n-C_0$.

Using (\ref{Wy1}) we obtain
\begin{equation}
\begin{array}{lcl}
W^2(x) - W'(x) &=&\displaystyle{ [ (B_1 - \sum_{i=1}^n g_i)^2 - C_0^{'} -
\sum_{i=1}^n (g_i^2 - 1)+1]/z^2 + [2(B_1-\sum_{i=1}^n g_i)C_0^{'}}\\
&&\displaystyle{+ 2(B_1 - \sum_{i=1}^n g_i)(\sum_{i=1}^n (g_i^2 - 1)) - (B_1 -
\sum_{i=1}^n g_i)+ 2 g_i(g_i^2 - 1)]/z}\\
&&\displaystyle{-\frac{3}{4z^4} + \sum_{i=1}^n \frac{1}{1+g_i z}[-2(B_1
- \sum_{i=0}^n g_i)(g_i^2-1)g_i - g_i^2(g_i^2-1)}\\
&&\displaystyle{+ 2C_0^{'}(g_i^2 - 1) - (g_i^4 -1) + \sum_{j\neq i}
\frac{(g_j^2 - 1) (g_i^2 - 1) g_i}{g_i - g_j}] + (C_0^{'})^2\ .
\label{iden}}\\
\end{array}
\end{equation}

We now make the following identification
\begin{equation}
W^2(x) - W'(x)  = V_1^{(DKV)}(x) - E\ ,
\end{equation}
where $E$ is the energy of the states in potential $V_1^{(DKV)}(x)$.

Then it follows that
\begin{equation}
-2(B_1 - \sum_{i=1}^n g_i)g_i - g_i^2 + 2C_0^{'} - (g_i^2 + 1) + 2
\sum_{i \neq j}
\frac{(g_j^2 - 1)g_i}{g_i - g_j} = 0\ ,
\label{eq1}
\end{equation}
\begin{equation}
2(B_1-\sum_{i=1}^n g_i)C_0^{'}+ 2(B_1 - \sum_{i=1}^n g_i)(\sum_{i=1}^n
(g_i^2 - 1) - (B_1 - \sum_{i=1}^n g_i) + 2\sum_{i=1}^n g_i (g_i^2 - 1)
 = -B\ ,
\label{eq2}
\end{equation}
\begin{equation}
(B_1-\sum_{i=1}^n g_i)^2 - C_0^{'} - \sum_{i=1}^n (g_i^2 - 1) + 1 = A\ ,
\label{eq3}
\end{equation}
\begin{equation}
(C_0^{'})^2  = -E \ .
\label{eq4}
\end{equation}
Multiplying (\ref{eq1}) by $g_i$ and summing over $i$ we obtain
\begin{equation}
-2(B_1-\sum_{i=1}^n g_i)\sum_{i=1}^n g_i - 2\sum_{i=1}^n g_i^3 +
2\sum_{i=1}^n C_0^{'} g_i - (2n-1) \sum_{i=1}^n g_i = 0\ .
\label{eq5}
\end{equation}
From (\ref{eq2}) and (\ref{eq5}) we get
\begin{equation}
B_1 = \frac{B}{1+2C_0} = \frac{B}{1 + 2\epsilon_0^{1/2}}\ .
\label{eq6}
\end{equation}
It can be verified by insertion that the wave functions
$\psi_n^{(-)}(x) = N \exp[-\int W(x) dx]$ are normalizable.
Equations (\ref{eq1}) and (\ref{eq3}) also imply that
\begin{equation}
A = \frac{(B/2)^2}{(n+\frac{1}{2}+\epsilon_n^{1/2})^2} + n^2 + n +1 +
(2n+1)\epsilon_n^{1/2}\ ,
\label{rel}
\end{equation}
where in obtaining the above relation we have taken $C_0^{'}=
-\epsilon_n^{1/2}$. We can summarize that our supersymmetric
construction reproduces exactly the results obtained in Ref.
\cite{Dutt}. Once we take $a=\epsilon_n^{1/2}$ it proves
equivalent to Eq. (\ref{eu}) of our preceding subsection
\ref{ni-ichi}.

\subsection{The allowed bound-state energies}
\label{ni-san}

Let us now continue with the analysis of the energy eigenvalues
based on the formula (\ref{eu}) and its equivalent form
(\ref{rel}) obtained in two different ways. The key element of our
approach is the strict observation of the constraints imposed on
the parameters by the boundary conditions of the wave functions.
By this we mean both the solutions of the ``old'' potential
$U_1(r)$ (\ref{u1pot}) and those of the ``new'' one
$V_1^{(DKV)}(x)$ (\ref{trip}).

The appropriate physical boundary condition for (\ref{u1wf})
near the threshold $r\to  0$ is standard, though a bit counterintuitive
\cite{comment,Landau}. Its implementation implies that we have to
choose $a > 1/2$. Then, after the transition from $r$ to $x$ we
get the wave functions still safely normalizable near the left
infinity $x \to -\infty$. Similarly, our explicit wave functions
remain asymptotically normalizable near the right infinities $r \to
\infty$ and $x \to +\infty$ if and only if we have $a+n <
b/(a+n)$. This means that the eligible quantum numbers $n = 0, 1,
\ldots, M$ have to be such that $0 \leq M < b^{1/2}-a$, i.e.,
 \begin{equation}
 (n+1/2)^2 < (a+n)^2 < b.
 \label{condeu}
 \end{equation}
As mentioned in Sec. \ref{ni-ichi},
for transition to the ``new" potential $V_1^{(DKV)}(x)$ we have
to re-parametrize $g_1=-B \equiv - 2b$ and
define the ``new" CES energy in terms of the ``old" ES coupling,
$k=a-1/2>0$. The second CES coupling $A > 2(n+1/2)^2+3/4$ is then
defined by (\ref{eu}), which is equivalent to (\ref{rel}) in Sec.
\ref{ni-ni}.
The $n-$dependence of the ``new" energy $a=a(n)=k_n+1/2
> 1/2$ is fully consistent with the $n-$independence of the
coupling $A$. For each level the CES potential $V(x)$ is a map of
a {\em different} ES potential $U(r)$. The energies are determined
by the cubic algebraic equation. In order to make this definition
unique we have to tell which one of the three roots of Eq.
(\ref{eu}) is ``physical". In Ref. \cite{Dutt} we find an advice
that ``from the three roots we can discard two by demanding that
the spectrum must reduce to the standard one for $B=0$". Such a
vague recipe is misleading since it is in manifest contradiction
with the above normalizability condition (\ref{condeu}) which
implies that $b=B/2 > 1/4$ cannot lie too close to zero.

\begin{figure}[h]
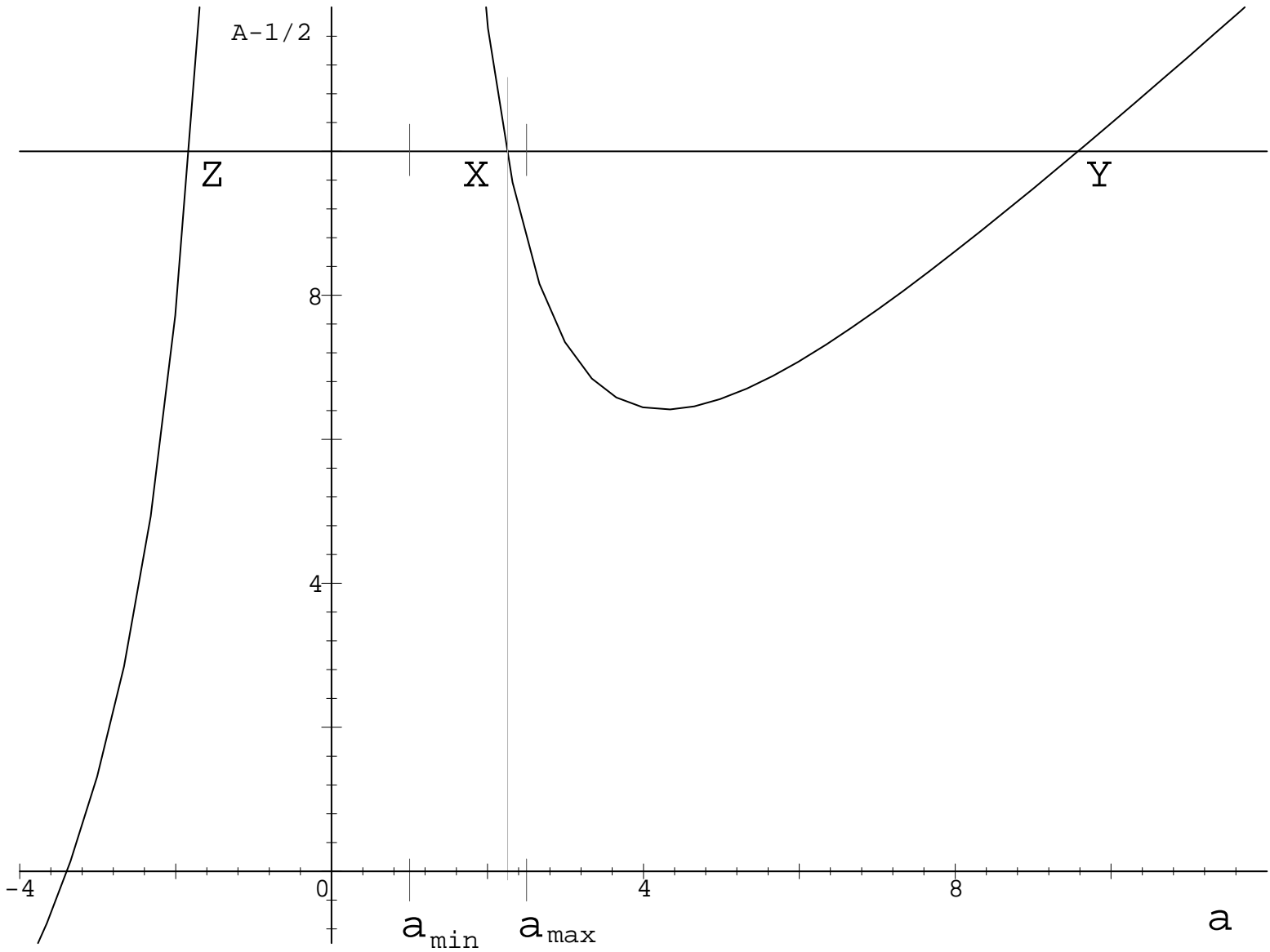

\caption{Graphical solution of Eq. (\ref{eu}).}
\label{fig1}
\end{figure}

The problem is not too difficult to disentangle. Equation
(\ref{eu}) has very transparent graphical interpretation in terms
of the intersection of the left-hand side horizontal line with the
right-hand side curve with three branches. The latter shape is a
sum of a growing linear term with a spike oriented upwards. Figure
\ref{fig1} indicates how one gets a triplet of roots in the $n=0$
ground state at $b=6.25$ and $A=10.25$. Always, only one of them
is compatible with the normalizability condition (\ref{condeu})
and lies in the ``admissible" interval $(0.5,2.5)$.

The general rule is that we always have to pick up the middle root
as physical. Let us give a proof of this assertion. Firstly we
re-name $b=\beta^2$ and re-scale our three roots, $Z =
[a_1(n)+n]/\beta$ $ < $  $X = [a_2(n)+n]/\beta$ $ < $ $Y =
[a_3(n)+n]/\beta$.
As long as $a(n) \in (1/2,\beta-n)$ we may
re-write Eq. (\ref{eu}) in the significantly simplified form
 \[
 \tau =
 \mu\,X + \frac{1}{X^2}
 \]
(and, similarly for $Y$ and $Z$) with abbreviations
 \[
 \tau = \tau (A,b,n) \equiv
 \frac{A+n^2+n-1/2}{b},
\ \ \ \ \ \ \ \ \  \ \ \ \ 0 < \mu=
 \frac{2n+1}{\beta}\,<2\ .
 \]
The leftmost root $Z$ will be always negative and can be discarded
immediately. Knowing that the acceptable root $X$ is constrained,
$X \in (T, 1)$, $T = (n+1/2)/\beta \in (0, 1)$,  it is now
sufficient to prove that the third root $Y$ {\em always} violates
our condition (\ref{condeu}),
 \begin{equation}
 [X \in (T,1) \ \& \ T \in (0,1)] \ \Longrightarrow \
 Y > 1.
 \label{trid}
 \end{equation}
For this purpose we eliminate $\tau$ and get the quadratic
equation
 \[
 \mu = \frac{X+Y}{X^2Y^2}.
 \]
We can skip the negative alternative and have the unique
definition of the root $Y$,
 \[
 Y=\frac{1+(1+4\mu\,X^3)^{1/2}}{2\,\mu\,X^2}.
 \]
As a smooth function of $\mu \in (0,2)$ and $X \in (T,1)$ it
satisfies our rule (\ref{trid}) {\em everywhere} within a
two-dimensional domain containing all points with $\mu<1$ and {\em
not containing} any point of the sign-changing boundary. This is
demonstrated quite easily. The boundary curve can be implicitly
defined as a set $X=\xi(\mu)$,
 \[
 {1+(1+4\mu\,\xi^3)^{1/2}}={2\,\mu\,\xi^2}\ .
 \]
Only on it the sign of $Y-1$ can change. This set is a part of the
curve defined by the square of the latter equation,
 \[
 \mu\,\xi^2 = \xi+1.
 \]
In the graphical language it is trivial to find that for the
positive $\xi > 0$ the right-hand side straight line intersects
the left-hand side parabola in a point which is a decreasing
function of $\mu$. Hence, the curve touches the boundary of our
open simplex of normalizability (with $\mu \in (0,2)$ and $X < 1$)
in a single point $(\mu=2, \xi=1)$. QED.

\section{Interpretation of the potential}
\label{san}

The potentials of Ref. \cite{Dutt} derived in two different ways
in Section \ref{ni} can be placed into a more general context by
realizing that both the point canonical transformation method \cite{pct}
presented in Sec. \ref{ni-ichi} and the supersymmetric construction
of Sec. \ref{ni-ni} can be formulated in terms of a rather general
approach based on the change of variables \cite{bs62,nat,lg89}.
In this section we specify these connections with the formulation of
Ref. \cite{lg89}, which can be considered a simplified treatment of
the general Natanzon-class potentials \cite{nat}.

Following the discussion of Ref. \cite{lg89} one considers the
Schr\"odinger equation
\begin{equation}
{{\rm d}^2 \psi \over {\rm d} x^2} +(E-V(x))\psi(x)=0
\label{sch}
\end{equation}
and assumes that its solutions can be written in the form
\begin{equation}
\psi(x)=f(x)F(z(x)) \ ,
\label{sol}
\end {equation}
where $F(z)$ satisfies a second-order differential equation
\begin{equation}
{{\rm d}^2 F \over {\rm d}z^2} +Q(z) {{\rm d} F \over {\rm d} z}
+R(z) F(z)=0 \ .
\label{specf}
\end{equation}
The function $F(z)$ can be any special function of mathematical
physics, e.g. the (confluent) hypergeometric function \cite{as70},
or any other function satisfying a second-order differential equation
of the type (\ref{specf}).
Simple calculation shows \cite{lg89} that the function $E-V(x)$ can
be written as
\begin{equation}
E-V(x)
=  {z'''(x) \over 2z'(x)} -{3 \over 4} \left({z''(x) \over
z'(x)}\right)^2 +(z'(x))^2\left[ R(z(x)) -{1 \over 2} {{\rm d}
Q(z) \over {\rm d}z} -{1 \over 4} Q^2(z(x))\right]\ ,
\label{emv}
\end{equation}
where the only unknown element is the function $z(x)$, which
basically governs the change of variables connecting the two differential
equations (\ref{sch}) and (\ref{specf}). Expressing $f(x)$ in (\ref{sol})
in terms of $z(x)$ and $Q(z)$, the solutions of the Schr\"odinger
equation can be written \cite{lg89} as
\begin{equation}
\psi(x)\sim (z'(x))^{-\frac{1}{2}}\exp\left(\frac{1}{2}
\int^{z(x)}Q(z){\rm d}z\right) F(z(x))\ .
\label{solfq}
\end{equation}
We are left with the task of finding such a functional form of
$z(x)$ which takes our Schr\"{o}dinger equation (\ref{emv}) into
an exactly and {\em completely} solvable problem.

Obviously, the transformation employed in Sec. \ref{ni-ichi}
(i.e. the point canonical transformation \cite{pct} or the Liouvillean
method \cite{Liouville}) is a special case of the above construction.
Taking
\begin{equation}
Q(z)=0 \hskip 3cm R(z)=-\kappa^2-U(z)\ ,
\label{subs}
\end{equation}
Eq. (\ref{emv}) reduces to the inverted version of Eq. (\ref{eu})
(with $r$ and $-k^2$ there replaced with $z$ and $E$ here). Similarly,
(\ref{solfq}) also reduces to the equivalent of (\ref{trenky}), where
$\chi(r)$ is playing the role of $F(z)$.

From here the approaches applied in Refs. \cite{bs62,lg89} and in
the point canonical transformation \cite{pct} emphasize somewhat
different strategies of deriving solvable potentials within the
Natanzon potential class \cite{nat}. In Refs. \cite{bs62,lg89} the
main point is to identify some term on the right-hand side of Eq.
(\ref{emv}), to account for the constant (i.e. the energy) term on
the left-hand side. With this, a differential equation of the type
\begin{equation}
\left(\frac{{\rm d} z}{{\rm d} x}\right)^{2}\phi(z)=C
\label{fifi}
\end{equation}
was obtained (see also \cite{lgmz00}), and this determined the function
$z(x)$ describing the variable transformation. In some cases
the $z(x)$ function could not be determined explicitly from (\ref{fifi}),
only the inverse $x(z)$ function, therefore a number of solvable models
obtained
this way turned out to be ``implicit'' potentials \cite{implicit,piii}.
On the other hand, following the point canonical transformation method
\cite{pct}, the $z(x)$ function is always available in an explicit
form, however, it is not guaranteed that any $z(x)$ function would lead
to a Schr\"odinger-like equation in which all the $n$-dependence can be
absorbed  into the constant (energy) term. Equation
(\ref{eu}) might turn out to have Sturm--Liouvillean form, where $n$
typically appears in coordinate-dependent terms. Simply stated, the
approach of Ref. \cite{lg89} focuses on having the energy in a simple
form, even on the expense of leaving the solutions in a complicated
(implicit) form, while in the point canonical transformation the
preference is having the solutions in an explicit form, rather than
getting the energy expression in a simple way. We stress that despite
this difference, the two approaches are interrelated, and are special
cases of deriving Natanzon-class potentials. We shall come back to this
point later on.

\subsection{Conventional construction}

Let us now see how potential $V_1^{(DKV)}(x)$ in Eq. (\ref{trip})
can be obtained from the method described in Ref. \cite{lg89}.
For this, $F(z)$ should be identified with a Jacobi polynomial:
$F(z)=P^{(\alpha,\beta)}_n(z)$. Equation (4.2) in
Ref. \cite{lg89} is an explicit form for $E-V(x)$ in this case:
\begin{eqnarray}
E-V(x)&=&
\frac{z'''(x)}{2 z'(x)}-\frac{3}{4}\left(\frac{z''(x)}{z'(x)}\right)^2
+\frac{(z'(x))^2}{1-z^2(x)}n(n+\alpha+\beta+1)
\nonumber\\
&&+\frac{(z'(x))^2}{(1-z^2(x))^2}\left[\frac{1}{2}(\alpha+\beta+2)
-\frac{1}{4}(\beta-\alpha)^2\right]
\nonumber\\
&&+\frac{(z'(x))^2 z(x)}{(1-z^2(x))^2}\frac{1}{2}(\beta-\alpha)(\beta+\alpha)
\nonumber\\
&&+\frac{(z'(x))^2 z^2(x)}{(1-z^2(x))^2}\left[\frac{1}{4}-
\left(\frac{\alpha+\beta+1}{2}\right)^2\right]\ .
\label{emvjac}
\end{eqnarray}
As discussed in Ref. \cite{lg89}, one selects differential equations
of the type (\ref{fifi}) for $z(x)$ to get constant terms on the
right-hand side of (\ref{emvjac}). In \cite{lg89} the first two non-trivial
terms were picked, leading to the PI and PII potential classes,
typical representations of which are, for example, $U_2(r)$ and $U_1(r)$
in Eqs. (\ref{u2pot}) and (\ref{u1pot}), respectively. The defining
differential equation of these is $(z')^2(1-z^2)^{-1}=C$ and
$(z')^2(1-z^2)^{-2}=C$.
Later in Ref. \cite{piii} the third ``PIII''
possibility,
$z(z')^2(1-z^2)^{-2}=C$ was also discussed, resulting in an
``implicit potential''. All these potentials are exactly solvable
Natanzon-class potentials, furthermore, those discussed in Ref.
\cite{lg89} also have the property of shape-invariance.

The fourth possibility,
\begin{equation}
z^2 (z')^2(1-z^2)^{-2}=C
\label{piv}
\end{equation}
was not discussed in detail in Ref. \cite{lg89}, only the generic form of
the solution was mentioned. However, it turns out, that the function
$z(x)=(1+\exp(2C^{1/2}x+D))^{1/2}$ satisfies (\ref{piv}), and it leads
to the same variable transformation as that discussed in Ref. \cite{Dutt},
if the $C^{1/2}=-1$ and $D=0$ choice is made. The actual form of
(\ref{emv}) is now (in the ``PIV'' case)
\begin{eqnarray}
E_n-V(x)&=&-\left(n+\frac{\alpha+\beta+1}{2}\right)^2
+\frac{1}{2}(\beta-\alpha)(\beta+\alpha) z^{-1}(x)
+\frac{3}{4}z^{-4}(x)
\nonumber\\
&&+\left[\left(n+\frac{\alpha+\beta+1}{2}\right)^2
-\left(\frac{\alpha+\beta}{2}\right)^2
-\frac{3}{4}-\frac{1}{4}(\beta-\alpha)^2\right]z^{-2}(x)\ .
\label{emvpiv}
\end{eqnarray}
This leads to a solvable potential if the $n$-dependence can be canceled
in the coordinate-dependent (i.e. potential) terms by a suitable change
of the parameters. Comparing (\ref{emvpiv}) with (\ref{trip}) we get
\begin{equation}
A=-\left[\left(n+\frac{\alpha+\beta+1}{2}\right)^2
-\left(\frac{\alpha+\beta}{2}\right)^2
-\frac{3}{4}-\frac{1}{4}(\beta-\alpha)^2\right]\ ,
\label{aaa}
\end{equation}
\begin{equation}
B=\frac{1}{2}(\beta-\alpha)(\beta+\alpha)\ ,
\label{bbb}
\end{equation}
and
\begin{equation}
E_n=-\left(n+\frac{\alpha+\beta+1}{2}\right)^2\ .
\label{piven}
\end{equation}
Obviously, $\alpha$ and $\beta$ depend on $n$ and also on the potential
parameters $A$ and $B$. Substituting (\ref{piven}) in (\ref{aaa}) and
combining it with (\ref{bbb}) we arrive at (\ref{eu}), the equation defining
the energy eigenvalues in the two approaches of Sec. \ref{ni}.

The bound-state wave functions are found to be
\begin{equation}
\psi(x)\sim z^{1/2}(x)(z(x)+1)^{\beta_n/2}(z(x)-1)^{\alpha_n/2}
P^{(\alpha_n,\beta_n)}_n(z(x))\ ,
\label{bswf}
\end{equation}
which (apart from some misprints), corresponds to Eqs. (15), (16) and (18)
in Ref. \cite{Dutt}, if we substitute $\alpha_n=B/(2c)-c$
and $\beta_n=-B/(2c)-c$.

\subsection{Supersymmetric connection}

In the knowledge of the bound-state wave functions, constructing
the superpotential $W(x)$ is a simple matter using Eq.
(\ref{sspot}). From (\ref{bswf}) with $n=0$ one obtains
\begin{equation}
W(x)=\frac{1}{2}(\alpha_0+\beta_0+1)+\frac{\alpha_0-\beta_0}{2z(x)}
-\frac{1}{2z^2(x)}\ .
\label{dkvw}
\end{equation}
In order to get closer to the methods described in Sec.
\ref{ni-ni}, we also introduce the singular superpotentials obtained
in a similar way from the wave functions with $n>0$. The Jacobi
polynomial appearing in these functions is best expressed in a
product form
\begin{equation}
P^{(\alpha_n,\beta_n)}_n(z)\sim\Pi_{i=1}^n (z-c_i)\ ,
\label{poly}
\end{equation}
where the $c_i$ are at the roots (nodes) of the polynomial. Obviously,
the logarithmic derivative of this product will reduce to a sum
form
\begin{equation}
\frac{\rm d}{{\rm d}x}(\ln P^{(\alpha_n,\beta_n)}_n(z))=
\frac{{\rm d}z}{{\rm d}x}
\frac{\rm d}{{\rm d}z}\sum_{i=1}^n \ln(z-c_i)
=(z^{-1}-z)\sum_{i=1}^n \frac{1}{z-c_i}\ .
\label{polysum}
\end{equation}
Here we used the differential equation (\ref{piv}) to express
$z'$ in terms of $z$. This explains the sum appearing in the
superpotential (\ref{Wy1}) in Sec. \ref{ni-ni}. A similar
construction can readily be presented for the superpotential used
in Ref. \cite{Nag} describing the potential of Ref. \cite{Dutra}
in a supersymmetric framework. The polynomial there is of the
Hermite type.

\subsection{Relation to the Natanzon potentials}

Our discussion in the present Section was based on the approach
of Ref. \cite{lg89}, which is general enough to incorporate both
the conventional and the supersymmetric formulation of potential
(\ref{trip}) in a relatively straightforward way. One can, however,
put the whole subject into an even more general framework, that
of the Natanzon potentials \cite{nat}. Although the discussion could
have been presented using the formalism of this potential class, we
decided to follow the easier route of Ref. \cite{lg89} for several
reasons. First, the general formalism was too heavy for demonstrative
purposes, and second, its relation to the machinery of supersymmetric
quantum mechanics \cite{Witten,Khare} is less transparent. However,
to conclude this Section we present the essential facts about
Natanzon potentials, and their relevance to the potentials we
investigated.

The general families of the Natanzon \cite{nat} and Natanzon confluent
\cite{natconf} potentials are characterized by the feature that their
solutions are expressed in terms of a single (confluent) hypergeometric
function. The general Natanzon potential depends on six parameters,
three of which ($f$, $h_0$ and $h_1$) appear explicitly in the
expression
\begin{equation}
V(x) =  -{z'''(x) \over 2z'(x)} +{3 \over 4}
\left({z''(x) \over z'(x)}\right)^2
+\frac{fz(x)(z(x)-1)+h_0(1-z(x)+h_1z(x)}{{\cal R}(z(x))}\ ,
\label{natpot}
\end{equation}
while three others ($a$, $c_0$ and $c_1$) enter implicitly through
the $z(x)$ function determined by the differential equation
\begin{equation}
z'(x)\equiv \frac{{\rm d}z}{{\rm d}x}=\frac{2z(1-z)}{({\cal R}(z))^{1/2}}
\label{natzv}
\end{equation}
with
\begin{equation}
{\cal R}(z)=a z(x)(z(x)-1)+c_0(1-z(x))+c_1 z(x)\ .
\label{natr}
\end{equation}
The construction of \cite{lg89}, when specified for the Jacobi
polynomials (a special case of the hypergeometric function \cite{as70})
can easily be recognized as a particular reformulation of this change
of variable method. (See also Ref. \cite{badh} and the Appendix of
Ref. \cite{lgmz00}.)
The energy spectrum is determined \cite{nat} by the implicit
equation
\begin{equation}
2n+1=(f+1-aE_n)^{1/2}-(h_0+1-c_0E_n)^{1/2}-(h_1+1-c_1E_n)^{1/2}\
\equiv \alpha_n-\beta_n-\delta_n\ ,
\label{naten}
\end{equation}
while the bound-state wave functions are written as
\begin{equation}
\psi(x)\sim {\cal R}^{1/4}(z(x))(1-z(x))^{\frac{\delta_n}{2}}
(z(x))^{\frac{\beta_n}{2}} F(-n,\alpha_n-n;\beta_n+1;z(x))\ .
\label{natwf}
\end{equation}
The form of (\ref{natwf}) is again reminiscent of the construction
of Ref. \cite{lg89}, while (\ref{naten}) is close to the implicit
energy formula obtained for the potential of Ref. \cite{Dutt} in
the point canonical transformation formalism.

Equations similar to those above are valid for the Natanzon confluent
potential class \cite{natconf} too.

It is instructive to examine the role of the 3+3 parameters appearing
in the Natanzon potentials, as it is related to the concept of
conditionally exact solvability. For the most commonly occuring
potentials (like the shape-invariant ones \cite{si}), the three parameters
determining the $z(x)$ function via (\ref{natzv}) and (\ref{natr}),
usually only one appears, and even that one is a trivial scaling
parameter of the coordinate and/or the energy scale. (Trivial
coordinate shifts can also appear through them.) Usually they play a
non-trivial role only in the case of some ``implicit'' potentials
\cite{implicit}.

The other three parameters appearing in (\ref{natpot}) set the potential
shape, and determine the relative strength of the individual potential
terms. In most potentials only one or two of these parameters
appear. The two parameters appearing in potential (\ref{trip}),
$A$ and $B$ are of this type. (There could be one more parameter
setting the length scale, but it is set to 1 in this case.)
Obviously, when there are three potential terms, as in (\ref{trip}),
and only two parameters, then the relative strength of the three
potential terms cannot be arbitrary, and has to be constrained.
This is why the third term of (\ref{trip}) is a numerical constant,
i.e. $-3/4$. It is the presence of this numerical constant which
earned potentials in Refs. \cite{Dutra,Dutt} the name ``conditionally
exactly solvable''. In fact, based on the structure of their
eigenfunctions, the potentials appearing in Ref. \cite{Dutt}
are of the Natanzon type \cite{nat}, while those in Ref. \cite{Dutra}
belong to the Natanzon confluent class \cite{natconf}. There are,
however, further considerations regarding normalizability and regularity,
which might impose restrictions on the solvability of certain potentials.
Not surprisingly, these may play a more important role in the case
of the less ``trivial'' potentials \cite{novaref}.

Finally, we note that the other class of CES potentials
\cite{susyces,pla} has completely different nature, and does not belong
to the Natanzon class, rather it has features typical for SUSY partners
of general Natanzon-class potentials. This again confirms our finding
that the concept of conditionally exact solvability is not an
alternative of exact solvability, rather it classifies potentials
according to different principles.

\section{Conclusions}

We analyzed the potentials introduced originally in Ref. \cite{Dutt} as
conditionally exactly solvable (CES) potentials via the method of
point canonical transformation. Our results concerned the following
three areas.

{\it i)} We gave a supersymmetric re-interpretation of this potential
class.

{\it ii)} We examined the cubic formula which determines implicitly
the energy eigenvalues of the problem. We rigorously took into account
boundary conditions of the eigenfunctions, and corrected certain
inaccuracies presented in Ref. \cite{Dutt}. We demonstrated that
from the three roots of the cubic equation there is only one (the
middle one) which can lead to physically acceptable eigenstates.

{\it iii)} We interpreted this potential in the general framework
of the Natanzon potential class, and demonstrated that this CES
potential, in fact, belongs to this class, and therefore it is
a {\it bona fide} exactly solvable problem.

\section*{Acknowledgements}

Partially supported by grant Nr. A 1048004 of the Grant Agency
of the Academy of Sciences of the Czech Republic and grant No.
T031945 of the OTKA (Hungary). G. L. acknowledges
the support of the J\'anos Bolyai Research Fellowship (Hungary).

\end{document}